\documentclass[useAMS,usenatbib]{mn2e}
\usepackage{graphicx}

\newcommand{\ms}{$\,$M$_\mathrm{\odot}$}
\newcommand{\ls}{$\,$L$_\mathrm{\odot}$}
\newcommand{\be}{\begin{equation}}
\newcommand{\ee}{\end{equation}}
\newcommand{\stars}{{\sc stars}}
\newcommand{\el}[2]{\ensuremath{^{#1}\mathrm{#2}}}

\title[Does simultaneous solution matter?]{Does simultaneous solution matter for stellar evolution codes?}
\author[R.~J. Stancliffe]{Richard J. Stancliffe\thanks{E-mail:
rs@ast.cam.ac.uk} \\
Institute of Astronomy, The Observatories, Madingley Road, Cambridge CB3 0HA, U.K. }
\begin{document}
\bibliographystyle{mn2e}

\date{Accepted 0000 December 00. Received 0000 December 00; in original form 0000 October 00}

\pagerange{\pageref{firstpage}--\pageref{lastpage}} \pubyear{0000}

\maketitle

\label{firstpage}

\begin{abstract}
A version of the \stars\ stellar evolution code has been developed that uses a non-simultaneous solution of the equations of stellar structure and evolution. In all other respects it is identical to the normal, fully simultaneous version. It is therefore possible to test the dependence of the solution on how the equations are solved.  Two cases are investigated: a 5\ms\ and a 3\ms\ star, both of metallicity Z=0.02. Prior to the asymptotic giant branch, the models are almost identical. However once thermal pulses start, the two methods of solution yield diverging results with the non-simultaneous technique predicting longer interpulse periods. This is traced to difficulties associated with hydrogen burning caused by the use of a moving mesh. It is shown that, with careful control of the temporal resolution, the results of the simultaneous technique can be recovered.
\end{abstract}
\begin{keywords}
methods: numerical, stars: evolution, stars: AGB and post-AGB
\end{keywords}

\section{Introduction}
Stellar evolution is based upon the solution of the equations of stellar structure, together with equations governing the mixing and burning of certain important isotopes. For a spherically symmetric star that is non-rotating, these equations are \citep[see e.g.][]{1990sse..book.....K}:
\begin{itemize}
\item{the equation of hydrostatic equilibrium,
\be
 {\mathrm{d}P \over \mathrm{d}m} = - {Gm \over 4\pi r^4},
\label{eq:hydrostatic}
\ee
where $P, r$ and $m$ are pressure, radius and the mass contained within a spherical shell of radius $r$ respectively.}
\item{The equation of mass conservation,
\be
 {\mathrm{d}r\over\mathrm{d}m} = {1\over4\pi r^2\rho},
\label{eq:massconservation}
\ee
where $\rho$ is density.}
\item{The equation of energy conservation,
\be
 {\mathrm{d}L\over\mathrm{d}m} = \epsilon,
\label{eq:energygeneration}
\ee
where $L$ is luminosity and $\epsilon$ is the energy generation rate including nuclear energy generation, energy from gravitational sources and energy losses from neutrino emission.}
\item{The equation of energy transport, which may be expressed as
\be
 {\mathrm{d}\ln T\over \mathrm{d}m} = - \bigtriangledown {\mathrm{d}\ln P\over\mathrm{d}m}
\label{eq:etransport}
\ee
where the form of $\bigtriangledown$ depends on whether the region of the star is radiative or convective.}
\end{itemize}
Assuming mixing is treated as a diffusive process\footnote{Not all codes treat mixing in this way.}, the equation governing mixing and burning for each isotope $X_i$ is given by:
\be
{\mathrm{d}\over\mathrm{d}m}\left(\sigma{\mathrm{d}X_i\over\mathrm{d}m}\right) = {\mathrm{d}X_i\over\mathrm{d}t} + R_{i} - S_{i},
\label{eq:composchange}
\ee
where $\sigma$ is the diffusion coefficient, $R_{i}$ is the rate at which the species $i$ is being burnt by nuclear reactions and $S_{i}$ is the rate at which it is being produced by nuclear reactions \citep{1972MNRAS.156..361E}.

While these equations may be common to all evolution codes, the method employed to solve them varies.  Three possible ways in which current codes can solve these equations may be identified. Before describing these methods it is necessary to define some terminology. A {\it timestep} is defined as the act of moving from a model at time $t$ to one at time $t+\Delta t$. To produce a model at a new timestep using the relaxation method it is necessary to make {\it iterations} on the solution. When the changes to the current solution are sufficiently small the model is said to have {\it converged}. Bearing in mind these definitions, three approaches to solving the equations may be defined: non-simultaneous, partially simultaneous and fully simultaneous.

The non-simultaneous approach involves converging a solution for the structure for a given timestep. This newly-computed structure is then used to calculate the mixing and burning for that timestep. Iterations are made separately on the structure and the chemistry (mixing and burning). Examples of codes that employ this method include those of \citet{2000MmSAI..71..719S} and \citet{2000A&A...360..952H}. The partially simultaneous approach involves solving for the structure equations for an iteration, then performing an iteration on the mixing/burning. One continues alternating between structure and chemistry iterations until the next timestep is converged. Such an approach is used in the Mount Stromlo Stellar Structure Program (MSSSP) employed by \citet{2002PASA...19..515K}, for example. 
Finally, the fully simultaneous approach involves solving all the equations together at each individual iteration of each timestep. This is the method employed by \citet{1971MNRAS.151..351E} in the original development of the \stars\ code and continues to be used in its current incarnations.

Current evolution codes predict different results for computations of the thermally pulsing asymptotic giant branch \citep{2003ApJ...586.1305L, 2004MmSAI..75..670S}. The reason for these discrepancies is uncertain. This paper addresses the effect the method of solution has on the results obtained. While it is thought that simultaneous solution will not influence the results of evolution calculations \citep{2005ARA&A..43..435H}, the effects have not been tested until now.

\section{The \stars\ code}
The stellar evolution code \stars, originally developed by \citet{1971MNRAS.151..351E} and most recently updated by \citet{1995MNRAS.274..964P} was used in this study. Derivatives of the original Eggleton code are unique in that they are the only codes to solve the equation of stellar structure and evolution in a fully simultaneous manner.

One would expect that if any difference exists between the methods of solution, it is likely to be greatest between the non-simultaneous and the fully simultaneous approaches. Therefore a version of the \stars\ code that employs a non-simultaneous method of solution has been developed. To do this, use has been made of a pair of subroutines designed to follow the evolution of minor isotopes \citep[see][for further details]{2005MNRAS.360..375S,stancliffe05}. These routines take the structure computed by the main routines and use it to compute the burning and mixing of material which is energetically unimportant. 

The version of the code employing a non-simultaneous method of solution works as follows. Equations \ref{eq:hydrostatic}-\ref{eq:etransport} are solved for using the normal evolution routines, with the mesh points being moved to their most appropriate locations by the mesh spacing function. It is therefore necessary to solve for the change in composition due to advection. Once a solution has been converged for the structure and the composition change due to mesh movement, equation \ref{eq:composchange} is then solved for each of the five important isotopes (\el{1}{H},\el{4}{He},\el{12}{C},\el{14}{N},\el{16}{O}) using the minor element subroutines, using the structure that has just been converged. The mesh spacing remains the same as for the evolution step and the diffusive mixing coefficient is taken from the converged structure. Thus it is possible to convert the method of solution employed by the \stars\ code into a non-simultaneous one by removing the equations of burning and mixing from the main code and placing them in these subroutines.

By changing the method of solution used by a code, we have two codes that are identical in all respects except for the one we are interested in. It is therefore possible to compare the effect that non-simultaneous solution {\it alone} has on the results obtained. This sort of comparison would be extremely difficult to do across different codes as different codes employ slightly different reaction rates, equations of state, mixing prescriptions, etc. and these would introduce uncertainties in to the source of any difference obtained. In addition there are different ways to write the difference equations used, different ways of averaging certain quantities \citep[see e.g.][]{2001MmSAI..72..299P} and different choices for defining variables at the centre or edges of shells. All these could affect the results obtained.  By creating a version of the \stars\ code that uses a non-simultaneous solution these problems are completely bypassed. In addition, the timestep control can be kept the same for both codes. 

The \stars\ code attempts to choose the most appropriate timestep size based on the changes to the variables of the previous model required to produce the current model. A sum of the absolute values of these changes (excluding those made to the luminosity) is made over all variables and over all meshpoints producing a single numerical value 
\[
 d = \Sigma_i \Sigma_k |\Delta x_{i,k}|,
\]
where $\Delta x_{i,k}$ are the values of the changes made to each variable $i$ at a given meshpoint $k$. This is then compared to a preset optimum value $d_\mathrm{opt}$. If $d_\mathrm{opt}/d$ is greater than one the timestep is increased by this fraction or 1.2, whichever is smaller. If $d_\mathrm{opt}/d$ is less than one the timestep is reduced by this fraction or 0.8, whichever is larger. In this way the most appropriate timestep is chosen for the next model.  The value of $d$ is dominated by the temperature and the degeneracy; the remaining variables make only minor contributions to the timestep control.

\section{Results}
To compare the differences between the two codes a 5\ms\ star of initial metallicity Z=0.02 was evolved from the pre-main sequence to the asymptotic giant branch without mass loss or convective overshooting using both simultaneous and non-simultaneous approaches. The initial model had 499 mesh points. Once second dredge-up was over, the models were remeshed with 1999 mesh points\footnote{This is twice the number normally employed and was used in case making the code non-simultaneous required additional spatial resolution.} and the AGB specific mesh spacing function of \citet*{2004MNRAS.352..984S} was employed in order to ensure proper spatial resolution of the important features of AGB stars.

Throughout the pre-main sequence, main sequence and red giant branch the evolution computed by the two methods is indistinguishable. At core helium burning the non-simultaneous solution displays a blue loop that extends toward the blue more than the simultaneous solution but this deviation is less than a tenth of a percent and should not be considered significant.

As the asymptotic giant branch is ascended the simultaneous solution grows a larger core than the non-simultaneous one. Prior to second dredge-up the simultaneous solution has a hydrogen-exhausted core that is about 0.3\% larger than the core of the non-simultaneous solution. The simultaneous solution then undergoes deeper second dredge-up than the non-simultaneous solution, resulting in it having a core that is 0.6\% less massive. This difference is about an order of magnitude smaller than the difference between the simultaneous solution and the comparable model of \citet{2002PASA...19..515K}, computed with a code employing a partially simultaneous method of solution. It therefore seems unlikely that the differences in details such as the core mass at first thermal pulse are due to the method of solution.

\begin{figure}
\includegraphics[width=7.5cm]{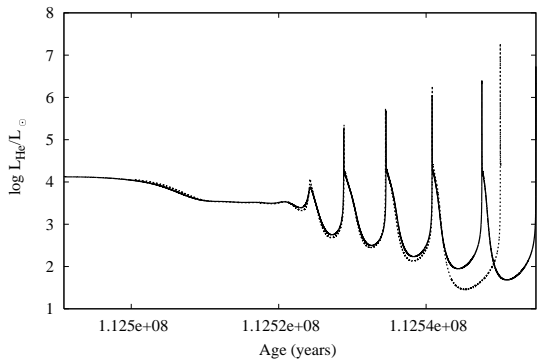}
\caption{Evolution of the helium luminosity as a function of age. The simultaneous solution is shown as a solid line; the non-simultaneous solution is the dashed line. The non-simultaneous solution has been shifted forward by $4.44\times10^4$\,yr to make the first thermal pulses coincident.}
\label{fig:baselhe}
\end{figure}

At the beginning of the thermally pulsing asymptotic giant branch (TP-AGB) the two models retain their similarities for a few thermal pulses. A rapid divergence in behaviour then sets in. This is demonstrated in the evolution of the helium luminosities for both models as shown in Figure~\ref{fig:baselhe}. The first three thermal pulses all have approximately similar peak helium luminosities. However, the interpulse period following the third thermal pulse is much longer for the non-simultaneous solution. This then results in the fourth thermal pulse of that model being much stronger.

\begin{figure}
\includegraphics[width=7.5cm]{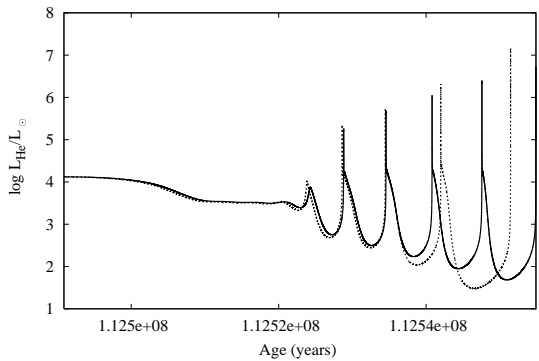}
\caption{Evolution of the helium luminosity as a function of age. The simultaneous solution is shown as a solid line; the non-simultaneous solution is the dashed line. Both sequences have been given the same inital model which was taken from just after second dredge-up.}
\label{fig:simstart}
\end{figure}

How much of the deviation is due to the fact that the two models have slightly different core mass at the beginning of the TP-AGB? While the initial difference is minimal it is not inconceivable that a small difference in initial conditions can lead to a larger difference later in the evolution. To assess the impact of just the difference in the method of solution, a starting model for a run using non-simultaneous solution was taken from the simultaneous solution run, just after second dredge-up. The evolution of the helium luminosity compared to the evolution from the simultaneous solution run is shown in Figure~\ref{fig:simstart}. Again, the initial pulses are similar but are seen to diverge after a couple of pulses. It would be desirable to continue the sequence to compare the behaviour over a longer run of thermal pulses but a numerical instability that could not be overcome in the non-simultaneous solution sequence meant this could not be done.

It therefore seems that simultaneous and non-simultaneous methods of solution do not yield the same results, despite the same timestep control routines being used in both cases.

\subsection{3\ms\ models}
Having noted that the two solution methods give different results, it is important to test whether these results extended to lower masses. Therefore 3\ms, Z=0.02 models were also produced using both codes. The details of the thermal pulse runs can be found in Tables~\ref{tab:recoupled} and \ref{tab:decoupled}.

\begin{table*}
\begin{center}
\begin{tabular}{cccccccc}
TP & M$_\mathrm{H}$ & $\tau_\mathrm{ip}$ & $\log$ (L$^{\mathrm{max}}_\mathrm{He}$/\ls) & $\Delta M_\mathrm{H}$ & $\Delta M_\mathrm{DUP}$ & $\lambda$ & C/O \\ 
   & (\ms) & ($10^4$\,yr) & & (\ms) & (\ms) & & \\
\hline
1 & 0.55267 &  ...  & 4.36792 & 0.01628 &  ...  &  ...  & 0.315 \\ 
2 & 0.56894 & 7.60 & 6.27639 & 0.00440 &  ...  &  ...  & 0.315 \\ 
3 & 0.57333 & 7.78 & 5.56767 & 0.00297 &  ...  &  ...  & 0.315 \\ 
4 & 0.57629 & 8.89 & 6.65021 & 0.00580 &  ...  &  ...  & 0.315 \\ 
5 & 0.58208 & 9.33 & 6.62750 & 0.00574 &  ...  &  ...  & 0.315 \\ 
6 & 0.58781 & 8.81 & 6.80665 & 0.00621 & 0.00047 & 0.076 & 0.317 \\ 
7 & 0.59355 & 8.22 & 6.90432 & 0.00633 & 0.00113 & 0.179 & 0.330 \\ 
8 & 0.59875 & 7.76 & 7.03324 & 0.00660 & 0.00190 & 0.288 & 0.356 \\ 
9 & 0.60345 & 7.50 & 7.16711 & 0.00701 & 0.00266 & 0.379 & 0.393 \\ 
\hline
\end{tabular}
\end{center}
\caption{Details of the 3\ms\ model computed using simultaneous solution. The data are TP -- the thermal pulse number, $M_\mathrm{H}$ -- the hydrogen free core mass, $\tau_{\mathrm{ip}}$ -- the interpulse period, L$^\mathrm{max}_\mathrm{He}$ -- the peak luminosity from helium burning, $\Delta M_\mathrm{H}$ -- the hydrogen free core mass growth during the interpulse, $\Delta M_\mathrm{DUP}$ -- the mass of material dredged up, $\lambda = \Delta M_\mathrm{DUP}/\Delta M_\mathrm{H}$ -- the dredge-up efficiency and C/O -- the surface carbon-to-oxygen ratio by number.}
\label{tab:recoupled}
\end{table*}

\begin{table*}
\begin{center}
\begin{tabular}{cccccccc}
TP & M$_\mathrm{H}$ & $\tau_\mathrm{ip}$ & $\log$ (L$^{\mathrm{max}}_\mathrm{He}$/\ls) & $\Delta M_\mathrm{H}$ & $\Delta M_\mathrm{DUP}$ & $\lambda$ & C/O \\ 
   & (\ms) & ($10^4$\,yr) & & (\ms) & (\ms) & & \\
\hline
1 & 0.54981 &  ...  & 5.64095 & 0.01704 &  ...  &  ...  & 0.315 \\ 
2 & 0.56684 & 6.87 & 5.33562 & 0.00241 &  ...  &  ...  & 0.315 \\ 
3 & 0.56925 & 8.99 & 6.45104 & 0.00517 &  ...  &  ...  & 0.315 \\ 
4 & 0.57441 & 10.99 & 6.65133 & 0.00591 &  ...  &  ...  & 0.315 \\ 
5 & 0.58032 & 10.59 & 6.87245 & 0.00642 &  ...  &  ...  & 0.315 \\ 
6 & 0.58666 & 10.01 & 6.96026 & 0.00663 & 0.00064 & 0.097 & 0.318 \\ 
7 & 0.59265 & 9.52 & 7.12672  & 0.00707 & 0.00142 & 0.201 & 0.328 \\ 
8 & 0.59830 & 9.17 & 7.28263 & 0.00755 & 0.00222 &  0.294 & 0.347 \\ 
\hline
\end{tabular}
\end{center}
\caption{Details of the 3\ms\ model computed using non-simultaneous solution. The headings are the same as in ~\ref{tab:recoupled}.}
\label{tab:decoupled}
\end{table*}

As with the 5\ms\ models, the 3\ms\ also display different behaviour, despite entering the TP-AGB with core masses within less than a percent of one another. The main difference
again seems to be in the interpulse periods which are again larger in the case of the non-simultaneous solution. The longer interpulse period also means that there is greater core growth in the non-simultaneous solution. The non-simultaneous solution also gives more violent thermal pulses. This is probably related to the longer interpulse period \citep[see the discussion in][]{2004MNRAS.352..984S}.

In terms of the onset of third dredge-up and its depth, both models give similar results. In both cases TDUP sets in when the H-exhausted core mass exceeds 0.585\ms\ and this happens on the sixth pulse in both models. Over the few thermal pulses with TDUP both models give comparable dredge-up efficiencies, with the simultaneous solution giving slightly lower efficiencies (which may be due to the slightly lower pulse strength). Curiously, while the simultaneous solution gives slightly less TDUP its C/O ratio becomes higher than in the case of the non-simultaneous solution. This occurs because the envelope in the case of the non-simultaneous solution goes deeper into the star during TDUP. This results in the carbon being mixed in with a greater amount of envelope material and hence becoming more diluted. 

\subsection{Testing TDUP}
Does the use of simultaneous solution affect the amount of TDUP predicted? It is difficult to assess the difference caused by using a simultaneous solution from the preceding runs because of the relationship between longer interpulse periods, more violent thermal pulses and deeper third dredge-up.

\begin{figure}
\includegraphics[width=7.5cm]{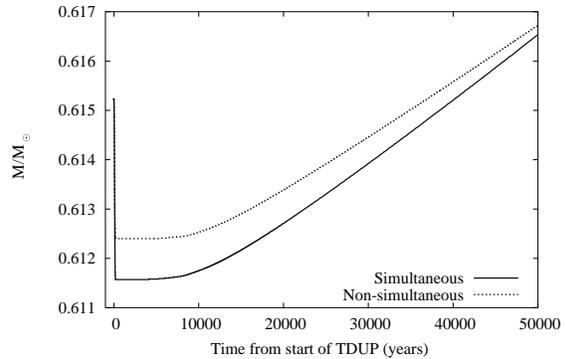}
\caption{Plot of the depth of TDUP following the 10th thermal pulse of the 3\ms\ sequence, using a starting model taken from the simultaneous-solution sequence. The run using simultaneous solution is the solid line; the dashed line is the non-simultaneous solution.}
\label{fig:TDUPtest}
\end{figure}

A sequence of models is therefore generated using both the simultaneous and non-simultaneous methods of solution, taking a starting model from just prior to the onset of TDUP after the 10$^\mathrm{th}$ thermal pulse of the simultaneous-solution run for the 3\ms\ star described above. The results are shown in Figure~\ref{fig:TDUPtest}. The effect is quite dramatic: the non-simultaneous solution gives much less TDUP. For the simultaneous solution 0.00366\ms\ of intershell material is dredged-up while for the non-simultaneous solution only 0.00283\ms\ is dredged-up. This corresponds to a dredge-up efficiency, $\lambda$, of 0.493 for the simultaneous solution versus 0.381 for the non-simultaneous solution.

\section{Discussion}
From the sample runs provided above it appears that there does exist a difference between the simultaneous and non-simultaneous solution of the equations of stellar structure and evolution. However, it appears that the important quantities of the core mass at first thermal pulse and the core mass at which third dredge-up occurs are not affected substantially by the use of a simultaneous or non-simultaneous solution as they differ by less than a percent in both these runs. By contrast, the difference between the core mass at first thermal pulse in the run using simultaneous solution is around 7\% smaller than the corresponding model of the code of \citet{2002PASA...19..515K}, which uses a partially simultaneous method of solution. This would suggest that the differences that occur between the predictions are not due to the method of solution used.

\begin{figure}
\includegraphics[width=7.5cm]{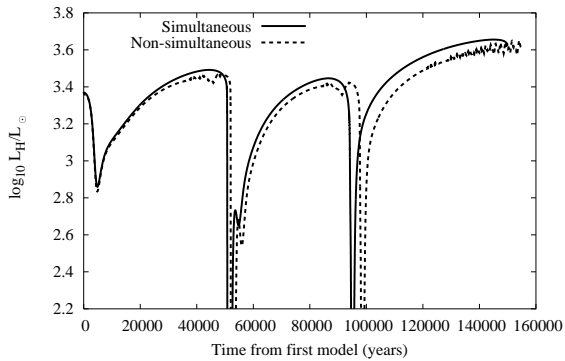}
\caption{Plot of the evolution of the hydrogen-burning luminosity as a function of time for a 1.5\ms\ model. The simultaneous solution is the solid line; the dashed line is the non-simultaneous solution. Note the erratic behaviour of the non-simultaneous solution prior to the drop in the H-luminosity at each thermal pulse.}
\label{fig:LH}
\end{figure}

What is the origin of the divergence in the model predictions presented above? The difference is first noticed in the interpulse period, which is when the H-burning shell is active. This is a logical place to start looking for a reason for the divergence. Figure~\ref{fig:LH} shows the evolution of the H-burning luminosity for a 1.5\ms\ Z=0.02 model, evolved from the same starting model with both codes. The non-simultaneous solution shows erratic behaviour in its hydrogen luminosity while the simultaneous solution is smooth as would be expected. 

\begin{figure}
\includegraphics[width=7.5cm]{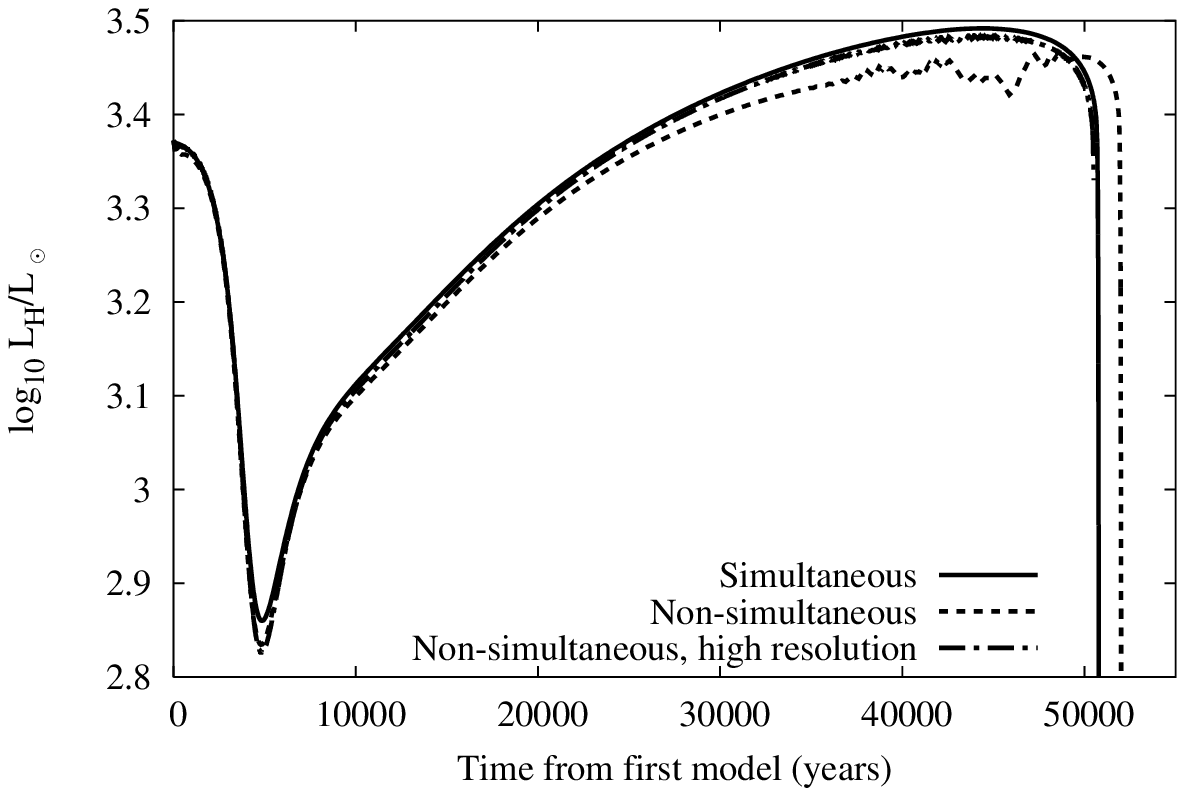}
\includegraphics[width=7.5cm]{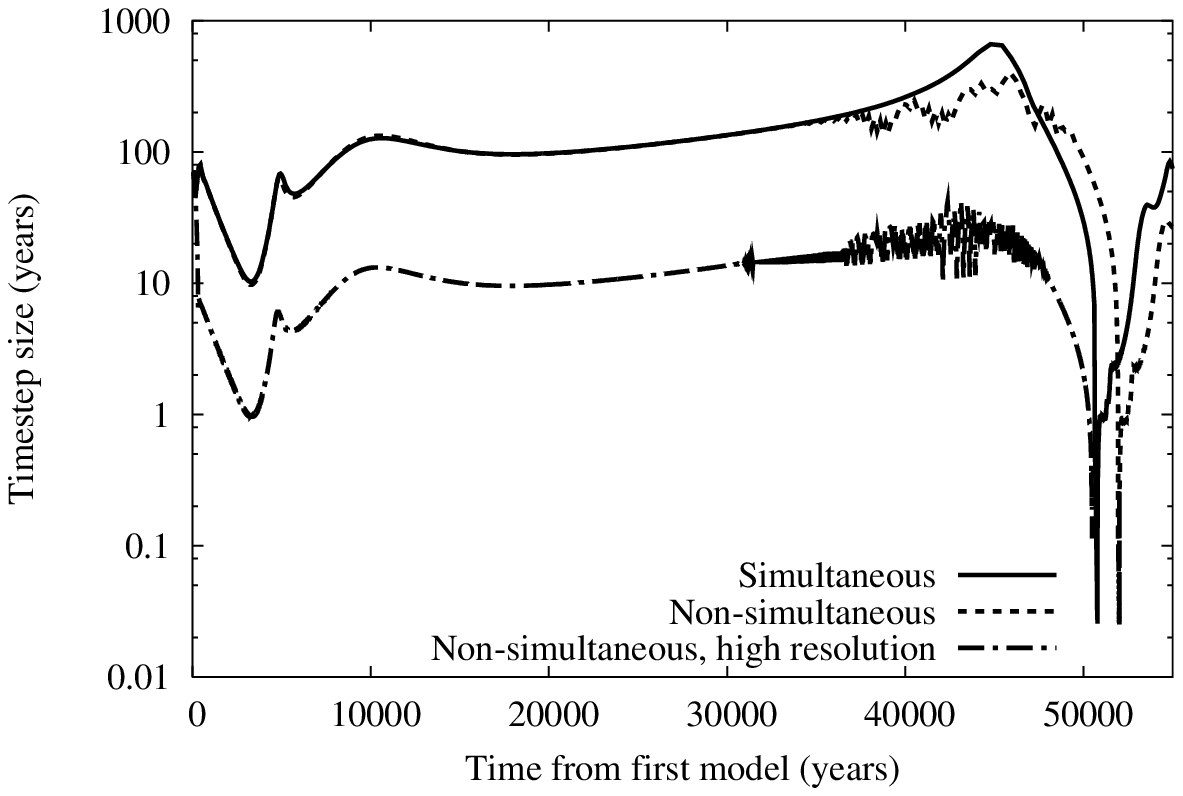}
\caption{Top panel: Plot of the evolution of the H-burning luminosity as a function of time for the 1.5\ms\ model. The solid line is the simultaneous solution, the dotted line is the non-simultaneous solution and the dot-dashed line is the non-simultaneous solution with higher temporal resolution. Bottom panel: Variation in timestep size as a function of time for the models in the top panel.}
\label{fig:timestep}
\end{figure}

The erratic behaviour of the H-burning luminosity in the non-simultaneous solution points to a resolution problem. By forcing the code employing the non-simultaneous solution to use smaller timesteps the simultaneous solution can be approached (see Figure~\ref{fig:timestep}). To obtain the high temporal resolution run the timestep has been reduced by at least an order of magnitude relative to the simultaneous solution run. The fluctuations in the timestep\footnote{They are caused by a timestep failing to converge. In such a case, the code tries again using a smaller timestep size.} around just prior to the thermal pulse suggest that there are still problems with the model even at these lower timesteps. A high spatial resolution model consisting of 4999 mesh points and using the same timestep control as the original non-simultaneous solution run gives the same erratic behaviour in the H-luminosity as the original run. This shows that the resolution problem is a temporal issue rather than a spatial one.

It therefore appears that a code employing a non-simultaneous method of solution can reproduce the results of one using a simultaneous solution, providing that more care is taken over the size of the timestep used. The bottom panel of Figure~\ref{fig:timestep} also shows that even at low timesteps the non-simultaneous solution still has difficulty with hydrogen burning as shown by the variation in timestep size around 40000 years from the initial model. Similar behaviour is found when the issue of TDUP is re-examined in this light. The non-simultaneous solution is seen to give comparable dredge-up to the non-simultaneous solution when the timestep size is reduced by over an order of magnitude.

\subsection{Lagrangian versus non-Lagrangian codes}

One caveat remains: the method used above to produce a code using a non-simultaneous method of solution does not produce one that is identical to the way most stellar evolution codes operate. Most codes in use are Lagrangian, i.e. the mesh points are placed at locations with fixed mass co-ordinates. This is not true of the \stars\ code. It was designed to use a non-Lagrangian, non-Eulerian mesh with a fixed number of mesh points; the mesh points are allowed to move in order to follow physically important features \citep{1971MNRAS.151..351E}. Eggleton's adaptive mesh (hereafter referred to as an Eggletonian mesh) has been retained in the version of the code employing a non-simultaneous method of solution.

\begin{figure}
\includegraphics[width=7.5cm]{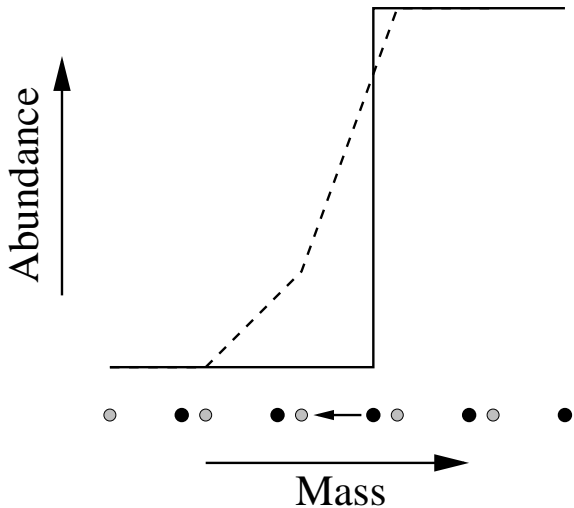}
\caption{A schematic depiction of why numerical diffusion occurs. A sharp abundance profile (solid line) is set up over a few mesh points (black dots). If the mesh points then move in mass to the positions indicated by the grey circles, the abundance profile is modified to that indicated by the dashed line.}
\label{fig:diffusionpic}
\end{figure}

It is possible that the combination of a non-simultaneous solution with an Eggletonian mesh is responsible for the instabilities outlined above. When the mesh moves during a structural step, the compositions are also changed to account for the movement. Where there are sharp gradients in composition this can cause problems. Suppose a profile with a sharp discontinuity is set up over a series of mesh points as depicted by the solid line in Figure~\ref{fig:diffusionpic}, with the black circles representing the location of the mesh points. If the mesh points move inward in mass, as indicated, to the locations marked by the grey circles, the profile is modified to that indicated by the dashed line. This is because when a mesh point moves the code works out the value of the abundance at the new location from the values from the old mesh point locations surrounding the new one. This inevitably leads to the sharp profile being smeared out. It is an unfortunate and very much undesirable consequence of using an adaptive mesh.

With this numerical diffusion, extra fuel can end up where it physically should not be. This results in the burning rate being increased. When the equations are solved, feedback to the structure can result and if it is strong enough it may create an instability. In a code employing simultaneous solution, this feedback is dealt with during the iterations required to converge a model so the problem does not arise. In a code employing a non-simultaneous solution, if the mesh moves slowly, or the timestep is low enough then the feedback between the structure and composition can be dealt with. The question is: how low must the timestep be for this to happen? As a na\"ive expectation, the timescale should be comparable to the time it would take for a given mesh point to traverse the mass difference between adjacent mesh points, i.e.:
\be
\tau_\mathrm{drift} = \left|{m_k - m_{k+1}\over \dot{m}_k}\right| 
\label{eq:drifttime}
\ee
where $m$ is the mass of the $k$th mesh point and $\dot{m}_k$ is the rate of change of mass of the $k$th mesh point due to the movement of the mesh (any given mesh point may move inward or outward in mass, depending on where the mesh spacing function thinks it should be). If the timestep of the code is smaller than this then changes should be properly resolved. In the above 1.5\ms\ model,  the drift timescale $\tau_\mathrm{drift}$ in the hydrogen burning shell is a little under 100 years and the timestep size is of this order. An instability is seen in the hydrogen burning luminosity. When smaller timesteps are used\footnote{This will reduce the magnitude of $\dot{m}$ and hence increase $\tau_\mathrm{drift}$.}, the timestep size is much less than the drift timescale and no instability is detected.

This problem would be avoided by using a Lagrangian mesh. From Equation~\ref{eq:drifttime} it can be seen that as $\dot{m}$ tends to zero (i.e. the mesh tends towards being Lagrangian), $\tau_\mathrm{drift}$ tends to infinity and the code will always have a timestep much smaller than the drift time. This could be tested by creating a Lagrangian code from the non-simultaneous code used here but this would be a major undertaking and is beyond the scope of this work.

However, it is possible to make a limited test of the situation. Using the viscous mesh technique of \citet{2004MNRAS.352..984S} it is possible to fix the mass locations of some of the mesh points. In this way, the regions of the mesh that are believed to be the problem (i.e. those mesh points from the surface to below the hydrogen burning shell) can be made to behave as Lagrangian mesh points. Unfortunately, when the mesh is locked it affects the timestep control mechanism typically producing lower timesteps than the unlocked mesh. Care must be taken to ensure that the effects of using a low timestep and the effects of using a Lagrangian mesh are not confused.

\begin{figure}
\includegraphics[width=7.5cm]{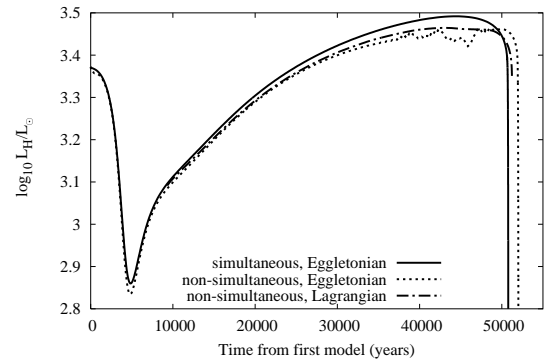}
\caption{Plot of the evolution of the hydrogen-burning luminosity as a function of time for the 1.5\ms\ model. The solid line represents the simultaneous solution which has an Eggletonian mesh, the dotted line represents the non-simultaneous code with the Eggletonian mesh and the dot-dashed line is the non-simultaneous code with a Lagrangian mesh near its surface and an Eggletonian mesh elsewhere. Note the absence of instabilities in the latter track.}
\label{fig:Lagrangian}
\end{figure}

The 1.5\ms\ model was re-run with the 700 mesh points closest to the surface being fixed in their mass co-ordinates. The base of the H-burning shell initially lies at around 500 mesh points from the surface so the problematic region is within the Lagrangian part of the mesh. Figure~\ref{fig:Lagrangian} shows the result. The H-luminosity evolves smoothly, though it does not replicate the behaviour of the simultaneous solution using the Eggletonian mesh. This is believed to be due to a lack of a resolution around the shell as a similar result is obtained if the upper 700 mesh points of the code employing the simultaneous solution are fixed.

It seems that if a non-simultaneous method of solution is to be used then a Lagrangian mesh is an advantage, as numerical diffusion in an Eggletonian mesh can lead to instability. However, the use of a fully simultaneous method of solution seems to avoid these problems. Further work on the effects of numerical diffusion within the framework of the \stars\ code needs to be done.

\section{Conclusions}
The dependence of the solution obtained on the method by which the equations of stellar structure and evolution are solved has been investigated for low and intermediate-mass stars. No difference was found prior to the AGB and the method of solution does not account for the differences seen in, for example, the core mass at first thermal pulse as computed by different evolution codes. 

Along the TP-AGB the two methods give diverging results, with the non-simultaneous method giving longer interpulse periods. This leads to more violent thermal pulses. The difference in interpulse period was tracked to an instability in the hydrogen burning shell of the non-simultaneous solution. This is due to numerical diffusion caused by the use of an Eggletonian mesh. By using smaller timesteps, or by using a Lagrangian mesh, the instability could be overcome.

The problems arising from the use of an Eggletonian mesh cloud the issue of a direct comparison between the non-simultaneous and simultaneous methods of solution. The discrepancies observed above are shown to be related to the use of an Eggletonian mesh together with a non-simultaneous method of solution. There is no evidence for an inherent problem with using a non-simultaneous method of solution. Further work needs to be done to assess the effect of using an Eggletonian mesh.

\section{Acknowledgements}
The author is grateful to the referee, John Lattanzio, for his careful reading of the manuscript and his many helpful suggestions that have added to the clarity of this work. RJS thanks Pierre Lesaffre for many useful discussions greatly improving mental, if not numerical, stability. He also thanks Churchill College for his fellowship.

\bibliography{stancliffe}

\label{lastpage}

\end{document}